\documentclass[a4paper,11pt]{article}
\pdfoutput=1
\usepackage{jcappub}

\usepackage{mathtools}

\newcommand{\rR}{\rho_R}
\newcommand{\rbh}{\rho_\text{BH}}
\newcommand{\gs}{g_\star}
\newcommand{\gss}{g_{\star s}}
\newcommand{\Teq}{T_\text{eq}}
\newcommand{\Min}{M_\text{in}}
\newcommand{\Tin}{T_\text{in}}
\newcommand{\Tev}{T_\text{ev}}
\newcommand{\Mbh}{M_\text{BH}}
\newcommand{\Tbh}{T_\text{BH}}
\newcommand{\mdm}{m_\text{dm}}
\newcommand{\Gsr}{\Gamma_\text{sr}}
\newcommand{\Ndmsr}{N_\text{dm}^\text{sr}}
\newcommand{\ndm}{n_\text{dm}}
\newcommand{\Trh}{T_\text{rh}}

\title{Superradiant Production of \\Heavy Dark Matter from \\Primordial Black Holes}

\author[a]{Nicolás Bernal,}
\author[b]{Yuber F.~Perez-Gonzalez}
\author[c]{and Yong Xu}
\affiliation[a]{Centro de Investigaciones, Universidad Antonio Nariño\\Carrera 3 este \# 47A-15, Bogotá, Colombia}
\affiliation[b]{Institute for Particle Physics Phenomenology, Durham University\\South Road, Durham DH1 3LE, United Kingdom}
\affiliation[c]{Bethe Center for Theoretical Physics and Physikalisches Institut, Universit\"at Bonn\\Nussallee 12, 53115 Bonn, Germany}

\emailAdd{nicolas.bernal@uan.edu.co}
\emailAdd{yuber.f.perez-gonzalez@durham.ac.uk}
\emailAdd{yongxu@th.physik.uni-bonn.de}

\abstract{Rotating black holes (BHs) can efficiently transfer energy to the surrounding environment via superradiance. In particular, when the Compton length of a particle is comparable to the gravitational radius of a BH, the particle's occupation number can be exponentially amplified. In this work, we investigate the effect of the primordial-black-hole (PBH) superradiant instabilities on the generation of heavy bosonic dark matter (DM) with mass above $\sim$ 1 TeV.
Additionally, we analyze its interplay with other purely gravitational and therefore unavoidable DM production mechanisms such as Hawking emission and the ultraviolet freeze-in.
We find that superradiance can significantly increase the DM density produced by PBHs with respect to the case that only considers Hawking emission, and hence lower initial PBH densities are required.}

\begin{document}
\begin{flushright}
    PI/UAN-2022-717FT\\
    IPPP/22/33
\end{flushright}

\maketitle

\section{Introduction}
Superradiance is a radiation enhancement phenomena, ubiquitous in many physical systems. In general relativity, a spinning black hole (BH) can feature superradiant instabilities, allowing efficient transfer of energy and angular momentum to the surrounding environment (for a review, see e.g. Ref.~\cite{Brito:2015oca}).
In particular, the frequency of a wave that scatters with a Kerr BH can be enhanced if it is smaller than the BH angular velocity~\cite{Teukolsky:1974yv}, similar to the Penrose process in the ergoregion of a Kerr BH~\cite{Penrose:1969pc}.
Additionally, Press and Teukolsky proposed the BH bomb Gedankenexperiment~\cite{Press:1972zz}: if one surrounds the BH with a reflecting mirror, the frequency can be further amplified when the wave bounces back and forth.
In reality, the gravitational interaction with the massive field that scatters can naturally act as a mirror~\cite{Cardoso:2004nk}.

For boson fields, the occupation number can be exponentially enhanced because of superradiant instabilities, leading to the formation of gravitational clouds around BHs, which could leave observable astrophysical signals. For example, these clouds could source gravitational waves (GWs) via $i)$ direct emission due to a dipole or quadrupolar configuration, $ii)$ through the transition between different harmonic quantum levels, and $iii)$ from boson condensate annihilation; for more details, see e.g. Refs.~\cite{Brito:2015oca, Brito:2021war}. This gives rise to a potential new avenue to constrain boson fields captured by superradiance.

There are two key parameters in superradiant instabilities: $i)$ the particle mass $\mdm$ (corresponding to a Compton length $\lambda_C = 1/\mdm$), and $ii)$ the BH mass $\Mbh$ (associated to a gravitational radius $r_g = G \, \Mbh$, $G$ being the gravitational constant). 
The detailed dynamics of the gravitational clouds invokes a solution of the Klein-Gordon equation in a Kerr metric.
Approximate solutions are reported in the literature in the limits $\alpha \gg 1$~\cite{Zouros:1979iw}, $\alpha \ll 1$~\cite{Detweiler:1980uk} and $\alpha \simeq 1$~\cite{Dolan:2007mj}, where $\alpha \equiv r_g/\lambda_C$ is the dimensionless ``fine-structure constant''.
Superradiance effects are maximized when $\alpha \simeq 1$~\cite{Dolan:2007mj}, corresponding to a boson mass $\mdm \simeq \mathcal{O}(10^{-11}) \times \left(M_{\odot} / \Mbh\right)$~eV~\cite{Brito:2021war}.
For astrophysical solar-mass BHs, it is then possible to probe ultralight bosons: see, e.g. the phenomenology of superradiance with QCD axions~\cite{Arvanitaki:2009fg, Arvanitaki:2014wva}, axion-like particles (ALPs)~\cite{Cardoso:2018tly, Stott:2018opm}, spin-1 vector fields~\cite{Endlich:2016jgc, Baryakhtar:2017ngi, Frolov:2018ezx, Cardoso:2018tly}, spin-2 fields~\cite{Brito:2013wya, Brito:2020lup}, and self-interacting fields~\cite{Baryakhtar:2020gao}.
Moreover, the effect of superradiant instabilities with even lighter bosons has recently been explored in the context of supermassive BHs~\cite{Davoudiasl:2019nlo, Davoudiasl:2021ijv, Chen:2022nbb}.

Complementary to previous studies that consider astrophysical BHs, in this work we instead investigate the phenomenology of superradiant instabilities in the context of {\it primordial} BHs (PBH), which could have been generated in the early Universe due to large density perturbations~\cite{Carr:2009jm, Carr:2020gox}.
PBHs with a large angular momentum could also have been formed, for example, $i)$ from collapsing domain walls~\cite{Eroshenko:2021sez}, $ii)$ during a matter-dominated era~\cite{Harada:2017fjm, Kuhnel:2019zbc}, or $iii)$ due to the assembly of matter-like objects such as Q-balls or oscillons~\cite{Flores:2021tmc}.
Alternatively, PBHs originally produced with a small angular momentum can gain spin\footnote{Note that here we mean the dimensionless spin parameter $a_\star$.} by emitting light scalar particles~\cite{Chambers:1997ai, Taylor:1998dk, Calza:2021czr}.
Here, we focus on PBHs with mass $\Mbh \lesssim \mathcal{O}(10^{9})~\text{g}$, which have fully evaporated before the onset of Big Bang nucleosynthesis (BBN), so that they would not jeopardize its successful predictions. During the evaporation process, any particle  including dark matter (DM) with mass small than the horizon temperature can be sourced~\cite{Hawking:1975vcx, Carr:1976zz}. If superradiant instabilities exist for supermassive astrophysical BHs, so do those BHs with smaller mass and primordial origin~\cite{Zouros:1979iw, Detweiler:1980uk}. Such instabilities with PBHs could naturally give rise to interesting cosmological signatures. For example, it has recently been shown that pion clouds can be formed around PBHs with mass around $\sim 10^{12}$~kg due to the superradiant instability~\cite{Ferraz:2020zgi}, offering a potential new avenue to constrain the abundance of PBHs by considering subsequent decay or annihilation of the pions.

In addition to studying the role of PBH superradiant instabilities in DM production, we pay particular attention to its interplay with other unavoidable gravitational channels such as UV freeze-in~\cite{Bernal:2018qlk, Bernal:2020ili, Bernal:2021akf}, and Hawking evaporation~\cite{Lennon:2017tqq, Hooper:2019gtx, Masina:2020xhk, Baldes:2020nuv, Gondolo:2020uqv, Bernal:2020kse, Bernal:2020ili, Bernal:2020bjf, Masina:2021zpu, JyotiDas:2021shi, Cheek:2021odj, Cheek:2021cfe, Sandick:2021gew}.
For illustration purposes, we first consider a spin-0 scalar field as the DM candidate; the extensions to the spin-1 and spin-2 cases will be investigated elsewhere.
We show that PBH superradiance gives rise to an unexplored nonthermal channel for generating heavy DM with mass above $\sim$ 1 TeV, and could source the entire observed abundance of DM relics.
We carefully track the evolution of DM by solving a system of coupled equations for the evolution of its number density, together with the PBH mass and spin. We find that sizable reductions on the initial PBH abundance are required in the regimes where superradiance is efficient.

The rest of the paper is organized as follows. We first briefly revisit the formation of and constraints on PBHs in Sect.~\ref{sec:PBH}.  Then in Sect.~\ref{sec:sr} we set up the Hawking evaporation formalism and then discuss superradiant instabilities with a particular focus on the evolution of the number of particles gravitationally bounded to PBHs.
In Sect.~\ref{sec:uv} we review the gravitational UV freeze-in. 
We collect the complete numerical results with all the previously-discussed gravitational contributions in Sect.~\ref{sec:full}. Finally, we summarize our findings in Sect.~\ref{sec:conclusion}.
Along the paper we denote the reduced Planck mass as $M_P \equiv 1 / (\sqrt{8\pi\, G})$, and we use natural units where $\hbar = c = k_{\rm B} = 1$.

\section{Primordial Black Holes} \label{sec:PBH}

Different mechanisms could lead to the formation of PBHs in the early Universe~\cite{Hawking:1975vcx, Carr:1974nx}.
For example, if above a threshold, density fluctuations can collapse into a BH when reentering the Hubble horizon~\cite{Villanueva-Domingo:2021spv}. 
Other scenarios for PBH formation include the collapse of cosmic string loops, the collapse of domain walls, or the collision of bubbles~\cite{Carr:2020xqk}.
Here, we remain agnostic about the formation mechanism, and assume that an initial PBH density was formed in a radiation-dominated era. 
For a plasma temperature $T = \Tin$, it is assumed that the initial mass of PBH $\Min$ is proportional to the horizon mass~\cite{Carr:2009jm, Carr:2020gox}, 
\begin{equation}\label{eq:Mi}
    \Min \equiv \Mbh(\Tin) = \frac{4\pi}{3}\, \kappa\, \frac{\rR(\Tin)}{H^3(\Tin)}\,,
\end{equation}
with numerical efficiency factor $\kappa \simeq 0.2$. Here $\rR$ and $H$ denote the radiation energy density and the Hubble rate, respectively.
Moreover, we adopt a monochromatic mass spectrum such that all PBHs were generated with same mass.%
\footnote{Extended PBH mass functions arise naturally if PBHs are created from inflationary fluctuations or cosmological phase transitions (see, e.g., Refs.~\cite{Hawking:1982ga, Carr:1994ar, Garcia-Bellido:1996mdl, Yokoyama:1998xd, Niemeyer:1999ak, Drees:2011hb, Kohri:2012yw, Kuhnel:2015vtw, Deng:2017uwc, Bernal:2022swt}), and a log-normal distribution represents a good approximation for models with a symmetric peak in the power spectrum of curvature perturbations~\cite{Dolgov:1992pu, Green:2016xgy, Kannike:2017bxn}.}
The extension to more realistic nonmonochromatic mass distributions will be investigated elsewhere. We characterize the PBH initial energy density $\rbh(\Tin)$ through the $\beta$ parameter, defined by
\begin{equation}
    \beta \equiv \frac{\rbh(\Tin)}{\rR(\Tin)}\,.
\end{equation}
Since PBHs behave as nonrelativitic matter, $\rbh$ redshifts slower than radiation, and a nonstandard cosmological expansion era could occur~\cite{Allahverdi:2020bys}.
A phase dominated by PBH with $\rbh > \rR$ can be triggered if the initial fraction is above a critical value $\beta_c$, 
\begin{equation}
    \beta_c \equiv \frac{\Tev}{\Tin}\,,
\end{equation}
where $\Tev$ is given by~\cite{Bernal:2021yyb, Bernal:2021bbv}
\begin{equation} \label{eq:Tev}
    \Tev \simeq \left(\frac{\gs(\Tin)}{640}\right)^{1/4} \left(\frac{M_P^5}{\Min^3}\right)^{1/2},
\end{equation}
and corresponds to the SM temperature at which the PBHs have completely evaporated. Here $\gs(T)$ represents the number of relativistic degrees of freedom contributing to the SM radiation energy density.
Note that if $\beta > \beta_c$, PBHs would start to dominate the total energy density at a temperature $T = \Teq$, when the equality between SM radiation and the PBH energy densities $\rR(\Teq) = \rbh(\Teq)$ occurred. 
Such a radiation-PBH equality temperature is given by
\begin{equation}
    \Teq = \beta\, \Tin \left(\frac{\gss(\Tin)}{\gss(\Teq)}\right)^{1/3},
\end{equation}
where $\gss(T)$ is the number of relativistic degrees of freedom that contribute to the SM entropy density.

In the evaporation process, all particles with a mass smaller than the horizon temperature can be emitted. 
If such evaporation occurred during or after BBN, the large particle injection could spoil the standard prediction of the abundance of light elements.
To avoid this, we consider PBHs that have fully evaporated before the onset of BBN, that is, $\Tev > T_\text{BBN} \simeq 4$~MeV~\cite{Sarkar:1995dd, Kawasaki:2000en, Hannestad:2004px, DeBernardis:2008zz, deSalas:2015glj, Hasegawa:2019jsa}, which can be translated into an upper bound on the initial mass of PBHs
\begin{equation} \label{eq:BBN}
    \Min \lesssim \mathcal{O}\left(10^9\right) \text{g}\,,
\end{equation}
as implied by  Eq.~\eqref{eq:Tev}.
On the other hand, one can obtain a lower limit on $\Min$ taking into account the upper bound on inflationary scale $H_I$ from constraints on the inflationary tensor-to-scalar ratio. According to the latest Planck collaboration results, one has $H_I \leq 2.5 \times 10^{-5}~M_P$~\cite{Planck:2018jri}, implying 
\begin{equation} \label{eq:CMB}
    \Min\gtrsim 4\pi\,\kappa\,\frac{M_P^2}{H_I}
    \simeq 0.1~\text{g}\,.
\end{equation}
Finally, the GWs produced due to evaporation of PBH could also spoil BBN, as they will contribute to the effective number of relativistic species $\Delta N_{\rm eff}$~\cite{Inomata:2020lmk, Domenech:2020ssp}.
This can be avoided if~\cite{Domenech:2020ssp}
\begin{equation} \label{eq:GW}
    \beta \lesssim 1.1 \times 10^{-6} \left(\frac{\kappa}{0.2}\right)^{-\frac12} \left(\frac{\Min}{10^4~\text{g}}\right)^{-\frac{17}{24}}.
\end{equation}
After describing the generic limits from evaporation, let us describe in detail the properties of spinning PBHs. 
Since superradiant amplification takes place only for such BHs, we will focus on Kerr PBHs in what follows.

\section{Superradiance and Hawking Radiation} \label{sec:sr}
In this section, we first have a quick overview of Hawking radiation, then focus on the particle number growth by BHs due superradiant instabilities. After that we put things together, and describe the interplay between superradiance and Hawking evaporation.

\subsection{Hawking Radiation from Kerr Black Holes}\label{subsec:Haw_KPBH}
PBHs could have been produced with a large initial angular momentum~\cite{Harada:2017fjm, Kuhnel:2019zbc, Eroshenko:2021sez, Flores:2021tmc, Chambers:1997ai, Taylor:1998dk, Calza:2021czr}.
For simplicity, we consider a monochromatic spin distribution, where all the PBHs had the same initial spin, without specifying the initial mechanism that generated it.
A Kerr BH is characterized by its instantaneous mass $\Mbh$ and dimensionless spin parameter $a_\star \equiv 8\pi\, J\, M_P^2/\Mbh^2 \in [0,1]$, with $J$ denoting the total angular momentum of the BH. 

The line element of the Kerr metric in Boyer-Lindquist coordinates is given by~\cite{Detweiler:1980uk}
\begin{equation}\label{eq:kerr}
    ds^2  = -dt^2 + \frac{2\, r_g\, r}{\Sigma} \left(dt - a_\star\, r_g\, \sin^2\theta\, d\varphi\right)^2 + \frac{\Sigma}{\Delta}\, dr^2 + \Sigma\, d\theta^2 + \left(r^2 + a_\star^2\, r_g^2\right) \sin \theta^2\, d\varphi^2,
\end{equation}
where $t$ denotes cosmic time and $(r,\, \theta,\, \varphi)$ are spherical coordinates, $\Sigma \equiv r^2 + a_\star^2\, r_g^2\, \cos^2 \theta$, and $\Delta \equiv r^2 - 2\, r_g\, r + a_\star^2\, r_g^2$, with $r_g \equiv G\, \Mbh$ denoting the gravitational radius.
The event horizon $r_+$ and the Cauchy horizon $r_-$ correspond to the coordinate singularities $\Delta =0$, and are given by
\begin{equation} \label{eq:event horizon}
    r_{\pm} = r_g \left( 1 \pm \sqrt{1 -a_\star^2} \right).
\end{equation}
The region $r_+ \leq r \leq r_t$, with
\begin{equation}
    r_t \equiv r_g\, \left(1 + \sqrt{1 -a_\star^2 \cos^2 \theta}\right),
\end{equation} 
denotes the so-called ergosphere, in which particles can gain energy from spinning BHs, and corresponds to the surface where the argument of $dt$ vanishes.
Meanwhile, the horizon temperature for spinning BHs is given by
\begin{equation}
    \Tbh = \frac{2M_P^2}{\Mbh}\, \frac{\sqrt{1-a_\star^2}}{1+\sqrt{1-a_\star^2}}\,.
\end{equation}
As mentioned above, during the Hawking evaporation process, all particles with masses smaller than $\Tbh$ can be generated. Within a time interval $[t,\, t + dt]$ and a momentum interval $[p,\, p + dp]$, the production rate for particle species $i$ with spin $s_i$, mass $\mu_i$ and internal degrees of freedom $g_i$ is~\cite{Hawking:1975vcx, Cheek:2021odj}
\begin{equation} \label{eq:spectrum}
    \frac{d^{2} \mathcal{N}_{i}}{d p\, d t}=\frac{g_{i}}{2 \pi^{2}} \sum_{l=s_{i}} \sum_{m=-l}^{l} \frac{\sigma_{s_{i}}^{l m}\left(\Mbh, p, a_\star\right)}{\exp \left[\left(E_{i}-m\, \Omega\right) / T_{\text{BH}}\right]-(-1)^{2 s_{i}}}\, \frac{p^{3}}{E_{i}}\,,
\end{equation}
where $E_i(p) = \sqrt{p^2 + \mu_i^2}$ denotes the energy, $\sigma_{s_i}$ describes the absorption cross-section, and $l$ and $m$ correspond to the total and magnetic quantum numbers, respectively.
Finally, in Eq.~\eqref{eq:spectrum}, $\Omega$ is the BH horizon angular velocity, a crucial quantity for superradiance, which is given by
\begin{equation} \label{Eq:Omega}
    \Omega \equiv \frac{4\pi\, a_\star}{1+\sqrt{1-a_\star^{2}}}\, \frac{M_P^2}{\Mbh}\,.
\end{equation}
Summing over all possible particle species and integrating over the phase space the rate in Eq.~\eqref{eq:spectrum}, one has the following set of equations describing the time evolution of the BH mass and spin, without including the effect of superradiance~\cite{Page:1976df} 
\begin{subequations} \label{eq:eveqsKerr}
    \begin{align}
        \frac{d\Mbh}{dt} &= -\varepsilon\left(\Mbh, a_\star\right) \frac{M_P^{4}}{\Mbh^{2}}\,, \\
        \frac{d a_\star}{d t} &=-a_\star\left[\gamma\left(\Mbh, a_\star\right)-2 \varepsilon\left(\Mbh, a_\star\right)\right] \frac{M_P^{4}}{\Mbh^{3}}\,,
    \end{align}
\end{subequations}
where the evaporation functions $\varepsilon\left(\Mbh, a_\star\right) \equiv \sum_i g_{i}\, \varepsilon_{i}(z_i)$ and $\gamma\left(\Mbh\right) \equiv \sum_i g_{i}\, \gamma_{i}(z_i)$ are given by the dimensionless quantities
\begin{subequations} \label{eq:epgK}
    \begin{align}
        \varepsilon_{i}\left(z_{i}, a_\star\right)&=\frac{27}{2 \pi^{2}} \int_{z_{i}}^{\infty} \sum_{l, m} \frac{\psi_{s_{i}}^{l m}\left(x, a_\star\right)\left(x^{2}-z_{i}^{2}\right)}{\exp \left(x^{\prime} / 2 f\left(a_\star\right)\right)-(-1)^{2 s_{i}}}\,  x d x\,,\\
        \gamma_{i}\left(z_{i}, a_\star\right) &= 108 \int_{z_{i}}^{\infty} \sum_{l, m} \frac{m\, \psi_{s_{i}}^{l m}\left(x, a_\star\right)\left(x^{2}-z_{i}^{2}\right)}{\exp \left(x^{\prime} / 2 f\left(a_\star\right)\right)-(-1)^{2 s_{i}}}\,  d x\,,
    \end{align}
\end{subequations}
with $z_i = \mu_i / \Tbh$, $\psi_{s_{i}}(E,\mu) \equiv 64 \pi M_P^2 \sigma_{s_{i}}(E,\mu)/(27 \Mbh^2)$, $x^{\prime} = x - m\, \Omega'$, $x = \Mbh\, E_i/M_P^2$, $\Omega' = \Mbh\, \Omega/M_P^2$, and $f(a_\star) = \sqrt{1 - a_\star}/(1 + \sqrt{1 - a_\star})$.
We refer the reader to Refs.~\cite{Page:1976df, Cheek:2021cfe, Cheek:2021odj} for further details.
We next describe the superradiant instability for Kerr BHs.

\subsection{Superradiance in a Nutshell}
When the Compton length $\lambda_C = 1/\mdm$ of a boson with mass $\mdm$ is comparable to the BH gravitational radius, a superradiant instability can occur, leading to an enhancement of the particle occupation number and a formation of bound states.
Such bound states are generally referred to as \emph{gravitational atoms}, characterized by a ``fine-structure constant''~\cite{Arvanitaki:2014wva}
\begin{equation}
    \alpha \equiv \frac{r_g}{\lambda_C} = G\, \Mbh\, \mdm \simeq 0.38 \left(\frac{\Mbh}{10^7~{\rm g}}\right)\left(\frac{\mdm}{10^7~{\rm GeV}}\right).
\end{equation}
Similarly to the hydrogen atom, the energy spectrum is controlled by three quantum numbers, namely the principal, orbital, and magnetic quantum numbers $(n,\, l,\, m)$, and at leading order is given by~\cite{Arvanitaki:2014wva, Ferraz:2020zgi}
\begin{equation} \label{Eq:omega}
    \omega \simeq  \mdm \left(1-\frac{\alpha^2}{2n^2}\right).
\end{equation}
Generically, in order to allow superradiance to occur, the energy or frequency $\omega$ has to be smaller than the BH angular velocity. In other words, if the latter is smaller than the former, PBHs would be unable to further amplify $\omega$. 
Let us now state the superradiance condition in a more mathematical way.
At leading order $\omega \sim \mdm$ (cf. Eq.~\eqref{Eq:omega}), and $\Omega \sim (2\, r_g)^{-1}$ for $a_\star \simeq 1$ (cf. Eq.~\eqref{Eq:Omega}).
To allow superradiance to occur, $\omega  < m\, \Omega$, which translates into
\begin{equation} \label{srcondition} 
      \alpha < \frac12\,,
\end{equation}
for $m = l = 1$. As mentioned above, due to superradiance, the frequencies can be enhanced, which could also be understood as a growth of the particle occupation number. The growth rate is defined to be the imaginary part of the bound state frequency $\mathfrak{Im} (\omega)$.\footnote{Further details can be found in Appendix~\ref{sec:appendix}.} In terms of the horizon radius $r_+$, for the dominant unstable mode ($n = 2$ and $l = m = 1$) the growth rate $\Gsr$  can be approximated as~\cite{Detweiler:1980uk, Brito:2015oca, Baryakhtar:2017ngi, Ferraz:2020zgi}
\begin{equation} \label{eq:Gsr}
    \Gsr = \frac{\mdm}{24} \left(\frac{\Mbh\, \mdm}{8\pi\,M_P^2}\right)^8 \left(a_\star - 2\, \mdm\, r_+\right),
\end{equation}
where $r_+$ represents the event horizon as defined in Eq.~\eqref{eq:event horizon}. Note that in the limit $a_\star \to 0$, $r_+ \to 0$, and therefore the growth rate $\Gsr$ vanishes, as expected, since superradiant instabilities do not occur for Schwarzschild BHs. This approximation models well the behavior of $\Gsr$ for small values of $\alpha$, however, fails in the large $\alpha$ regime.
A better determination of $\Gsr$ requires a numerical solution.

\begin{figure}
    \def\sepf{0.60}
	\centering
    \includegraphics[scale=\sepf]{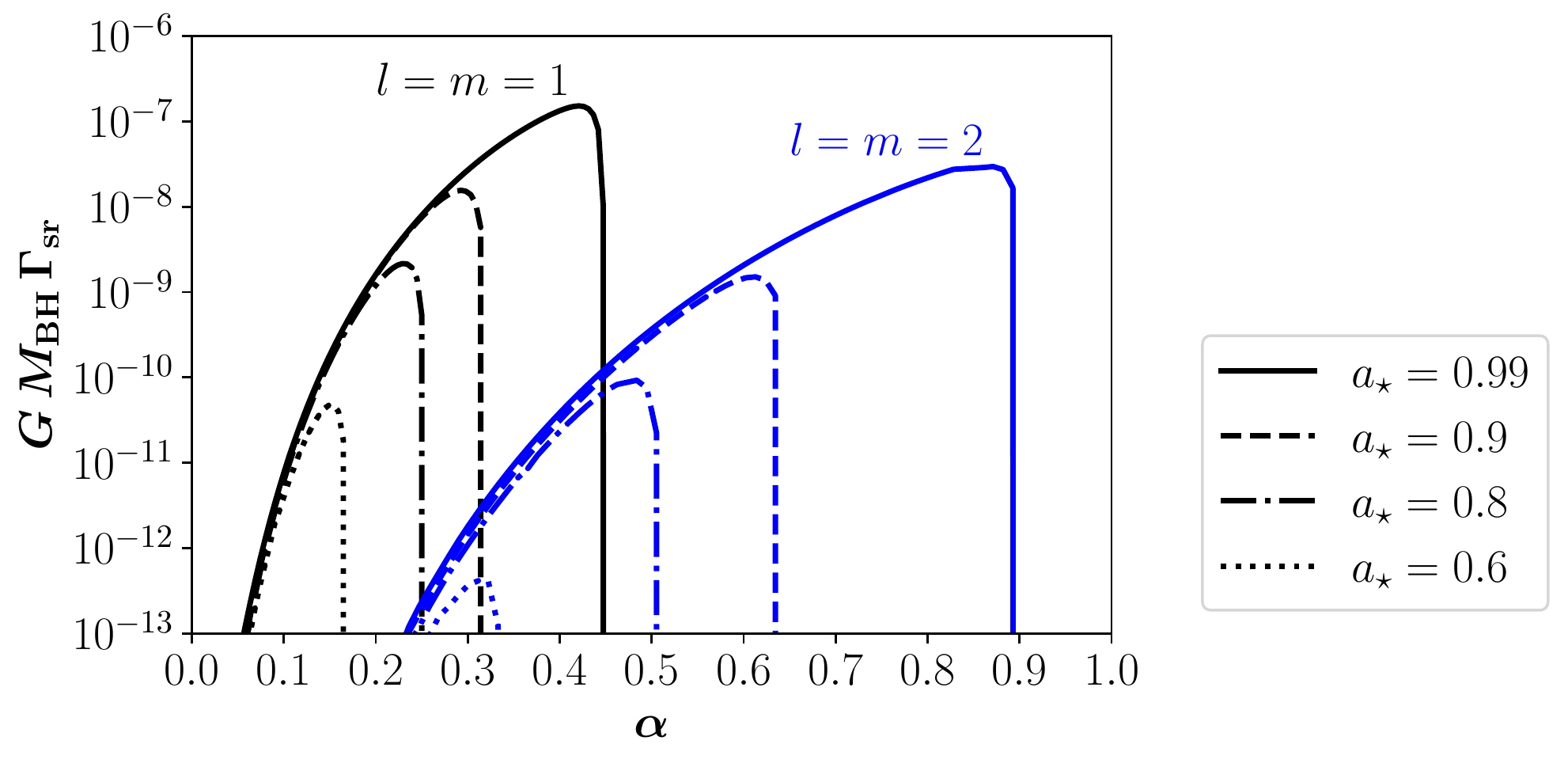}
    \caption{Normalized growth rate $\Gsr$ as function of the fine-structure constant $\alpha$, for $l = m = 1$ (black) or $l = m = 2$ (blue), and different values of the initial BH spin.
    }
	\label{fig:UV_sr_num}
\end{figure}

For completeness, we next describe the numerical method that we have used to compute the growth rates, following Ref.~\cite{Dolan:2007mj}.
To determine the frequencies of the bound states, it is required to solve the Klein-Gordon equation on a Kerr background~\cite{Dolan:2007mj, Cardoso:2018tly}.
After a standard variable separation, two power series solutions are proposed to solve the radial and angular equations, which lead to three-term recurrence relations between the different expansion coefficients~\cite{Leaver:1985ax, Dolan:2007mj}.
Implicit conditions for continued fractions, which are only satisfied by the bound-state frequencies, are obtained after performing algebraic manipulations.
Thus, one needs to employ numerical method to solve such implicit conditions and obtain the growth rate.
We use a {\tt scipy} standard root finder that applies the {\tt HYBR} method~\cite{2020SciPy-NMeth} to simultaneously solve the radial and angular implicit conditions. 
We have checked that our code gives the same rates as those presented in Ref.~\cite{Dolan:2007mj}, which we show in Fig.~\ref{fig:UV_sr_num} for $l = m = 1$ (black) and $l = m = 2$ (blue), for $a_\star = 0.99$, 0.9, 0.8 and 0.6.
In what follows, we only consider the instability on the state $l = m = 1$ since it grows exponentially faster than the others. 
Contributions to the DM abundance coming from other $l$ and $m$ states will be discussed elsewhere.
Detailed numerical simulations show that for $a_\star \simeq 1$, the superradiant instability is largest when $\alpha \sim 0.42$~\cite{Dolan:2007mj, Witek:2012tr}, as shown in Fig.~\ref{fig:UV_sr_num}.

Superradiant instabilities exponentially amplify the number of DM particles.
Defining $\Ndmsr$ to be the total number of DM particles gravitationally bounded to a PBH and considering for the moment only superradiance, the dynamics of the system is governed by the system of coupled equations~\cite{Arvanitaki:2014wva, Brito:2015oca, Roy:2021uye}
\begin{subequations} \label{eq:SR_Q1}
    \begin{align}
        \frac{d\Ndmsr}{dt} &= \Gsr\, \Ndmsr\,,\\
        \frac{d\Mbh}{dt} &= - \mdm\, \frac{d\Ndmsr}{dt}\,,\\
        \frac{dJ}{d t} &= - \sqrt{2}\, \frac{d\Ndmsr}{dt}\,.
    \end{align}
\end{subequations}
Apart from having $\alpha \sim 1$, superradiant exponential growth requires a nonvanishing initial population of gravitationally bounded particles.
However, it is interesting to realize that the exact number of initial particles has a small impact on its final number.
Noticing that
\begin{equation}
    \frac{d\Ndmsr }{d a_\star} = \frac{d \Ndmsr}{d t} \times \frac{dt}{d a_\star} =  \left[2\, \frac{\mdm}{\Mbh}\, a_\star - 8\pi \sqrt{2} \left(\frac{M_P}{\Mbh}\right)^2\right]^{-1} \simeq -\frac{1}{8\pi \sqrt{2}} \left( \frac{\Mbh}{M_P} \right)^2,
\end{equation}
it can be seen that the final amount of DM particles $\Ndmsr(\text{end})$ after the superradiant instability is
\begin{equation} \label{eq:Nend}
    \Ndmsr(\text{end}) \simeq \Ndmsr(\text{in}) + 1.5 \times 10^{19}  \left(\frac{\Mbh}{10^{5}~\text{g}}\right)^2 \left(a^{\text{ini}}_\star - a^{\text{end}}_\star\right) \simeq \mathcal{O}\left(10^{18}\right) \left(\frac{\Mbh}{10^{5}~\text{g}}\right)^2,
\end{equation}
featuring a small additive dependence on the initial number of DM particles $\Ndmsr(\text{in})$. 

\subsection{Interplay of Superradiance and Hawking Radiation}
\label{sec:sr_Haw}
As mentioned in Sect.~\ref{subsec:Haw_KPBH}, on top of the contribution from superradiance, DM particles are also unavoidably sourced by Hawking evaporation. In this section we investigate the interplay between the two mechanisms. The evolution of the system is governed by the set of coupled equations:
\begin{subequations} \label{eq:super_Hak}
    \begin{align} \label{eq:Ndmsr}
        \frac{d\Ndmsr}{dt} &= \Gsr\, \Ndmsr\,,\\
        \frac{d\Mbh}{dt} &= -\varepsilon\, \frac{M_P^4}{\Mbh^2} - \mdm\, \Gsr\, \Ndmsr\,,\\
        \frac{d a_\star}{d t} &= - a_\star \left[\gamma - 2 \varepsilon\right] \frac{M_P^4}{\Mbh^3} - 8 \pi \left[\sqrt{2} - 2\alpha\, a_\star\right]\Gsr\, \Ndmsr\, \frac{M_P^2}{\Mbh^2}\,,\\
        \frac{d\ndm}{dt} + 3 H\, \ndm &= n_\text{BH} \left[\Gamma_\text{BH$\to$DM} + \Gsr\, \Ndmsr\right],
    \end{align}
\end{subequations}
with $\ndm$ and $n_\text{BH}$ denoting the DM and PBH number density, respectively. The first line describes the usual evolution of total bounded DM numbers due to superradiant instability. The second and third lines are a combination of Eqs.~\eqref{eq:eveqsKerr} and~\eqref{eq:SR_Q1}, and describe the evolution of PBH mass and spin. Finally, the last line denotes the modified Boltzmann equation for the DM number density with contributions from both the superradiant instability and the Hawking evaporation. 
In the first term of the right-hand side (RHS), $\Gamma_\text{BH$\to$DM}$ denotes the DM production rate from BH through evaporation, which can be obtained by integrating Eq.~\eqref{eq:spectrum} over the phase space for a scalar DM field~\cite{Cheek:2021odj}. 
In the second term, $\Gsr\, \Ndmsr$ corresponds to the DM production rate per BH. Note that we are assuming that after the evaporation of the PBHs, DM behaves as free particles.
This allows us to explore the maximum reach of superradiant DM production.
The details the fate of the DM clouds will be explored in a future work.
Finally, Eqs.~\eqref{eq:super_Hak} reduce to Eqs.~\eqref{eq:eveqsKerr} or~\eqref{eq:SR_Q1} in the limit where superradiance or Hawking emission are artificially switched off. 

To allow superradiance to take place, the PBH should not evaporate before the instability becomes significant.
Thus, we can compare the typical timescale for superradiance to the lifetime of PBHs.
The PBH lifetime is roughly~\cite{Bernal:2020bjf}
\begin{equation}
    \tau =\frac{160}{\pi \gs} \frac{\Min^3}{M_P^4}\,,
\end{equation}
whereas the typical timescale for superradiance $t_{\text{sr}}$ corresponds to the inverse of the superradiance rate in Eq.~\eqref{eq:Gsr},
\begin{equation}
    t_\text{sr} \sim  \frac{1}{\Gsr} \simeq \frac{24}{\mdm}  \left(\frac{8\pi\,M_P^2}{\Min\, \mdm}\right)^8  \frac{1}{a_\star}\,.
\end{equation}
To develop superradiance, $t_{\text{sr}} <  \tau$ is required, so we can estimate a lower bound on the initial PBH mass, depending on the DM mass and initial PBH spin
\begin{equation} \label{eq:Mines}
    \Min \gtrsim 70 \left(\frac{1}{a_\star}\right)^{1/11} \left(\frac{10^{11}~\text{GeV}}{\mdm}\right)^{9/11} \text{g}\,.
\end{equation}
However, we stress that Kerr BHs lose angular momentum much faster than their mass~\cite{Page:1976df}, so they effectively become Schwarzschild BHs before completely evaporating.
As the BH should have enough angular momentum for the superradiant instability to rise, the estimate in Eq.~\eqref{eq:Mines} is conservative, since it does not consider the time when the BH would have shed their spin.
Still, we have verified that this characteristic of Kerr BHs only affects the estimated lower bound for light PBHs $\Min \lesssim 10$~g.

\begin{figure}[t!]
    \def\sepf{0.51}
	\centering
    \includegraphics[width=\textwidth]{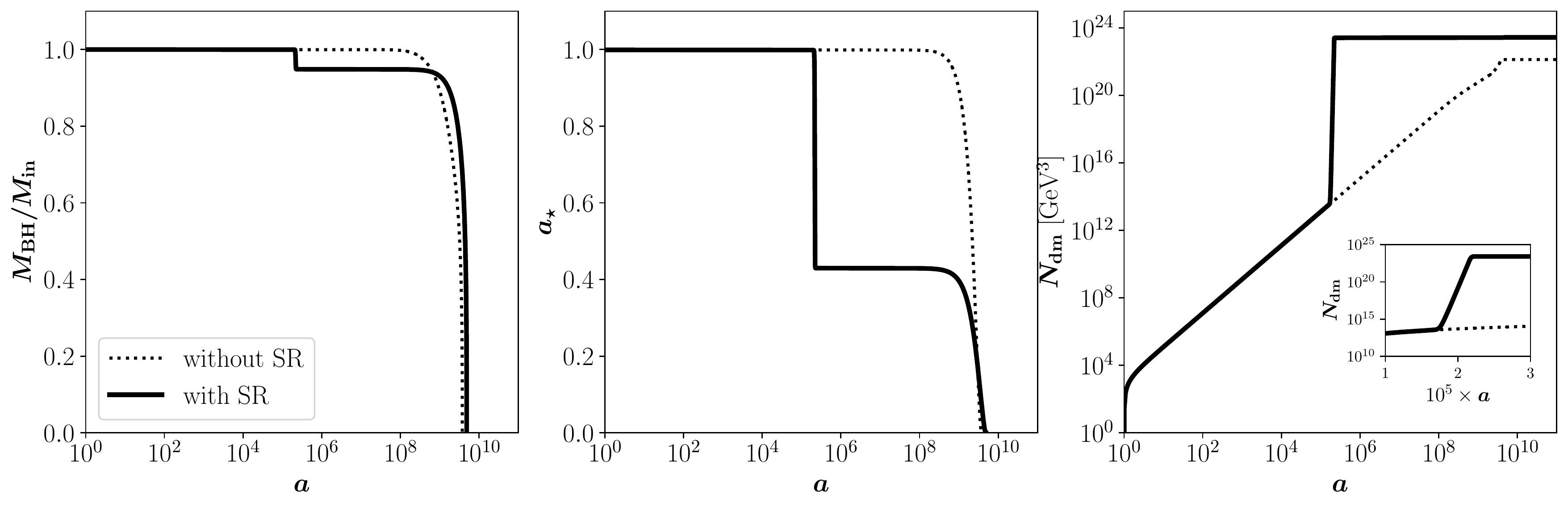}
    \caption{Evolution of the PBH mass (left), its spin (center), and the DM comoving number density (right) for $\Min = 3.2\times 10^4 $~g, $a_\star=0.999$, $\beta= 4.4 \times 10^{-22}$, and $\mdm = 10^9$~GeV.
    The dotted lines only take into account Hawking emission, whereas the solid lines also account for superradiance.
    The latter case reproduces the observed DM abundance. Note that SR represents superradiance.
    }
	\label{fig:solution}
\end{figure} 
The coupled system~\eqref{eq:super_Hak}, together with the Friedmann equations for the PBH and SM radiation energy densities are numerically solved using the Python package {\tt ULYSSES}~\cite{Granelli:2020pim}.%
\footnote{\url{https://github.com/earlyuniverse/ulysses}}
An example of the evolution of the PBH mass and spin, and the DM number density is shown in Fig.~\ref{fig:solution}, for $\Min = 3.2\times 10^4 $~g, $a_\star = 0.999$, $\beta = 4.4 \times 10^{-22}$ and  $\mdm = 10^9$~GeV, a benchmark that reproduces the observed relic abundance once superradiance is included.
The dotted lines only take into account Hawking emission, whereas the solid lines also contain the contribution from superradiance.
The effect of the superradiant instability manifests itself significantly when $t \sim 1/\Gsr$, which corresponds to $a/a_\text{ini} \sim \mathcal{O}\left(10^{5}\right)$ as depicted in Fig.~\ref{fig:solution}, with $a_\text{ini}$  being the scale factor at $T = \Tin$.
For simplicity, we define $a_\text{ini} = 1$ in the following.
At the end of the superradiant era, the BH mass is reduced only by a factor of $\sim 5\%$, while the spin decreases to $\sim 40\%$ of its initial value.
The right panel of Fig.~\ref{fig:solution} tracks the comoving DM number density, namely $N_\text{dm}=a^3 \ndm$. 
The superradiant instability exponentially amplifies the DM number density by $\sim\mathcal{O}(11)$ orders of magnitude in a short period of time.
Finally, once superradiance ends, the PBH evolution is governed by the usual Hawking evaporation.
Interestingly, compared to the case where superradiance is absent, the final DM number density gets boosted by factor of $\sim 21$. 
Hawking radiation produces an additional contribution to the total DM number density, but its impact is negligible compared to the abundance produced by superradiance. 

\begin{figure}[t!]
    \def\sepf{0.52}
	\centering
    \includegraphics[scale=\sepf]{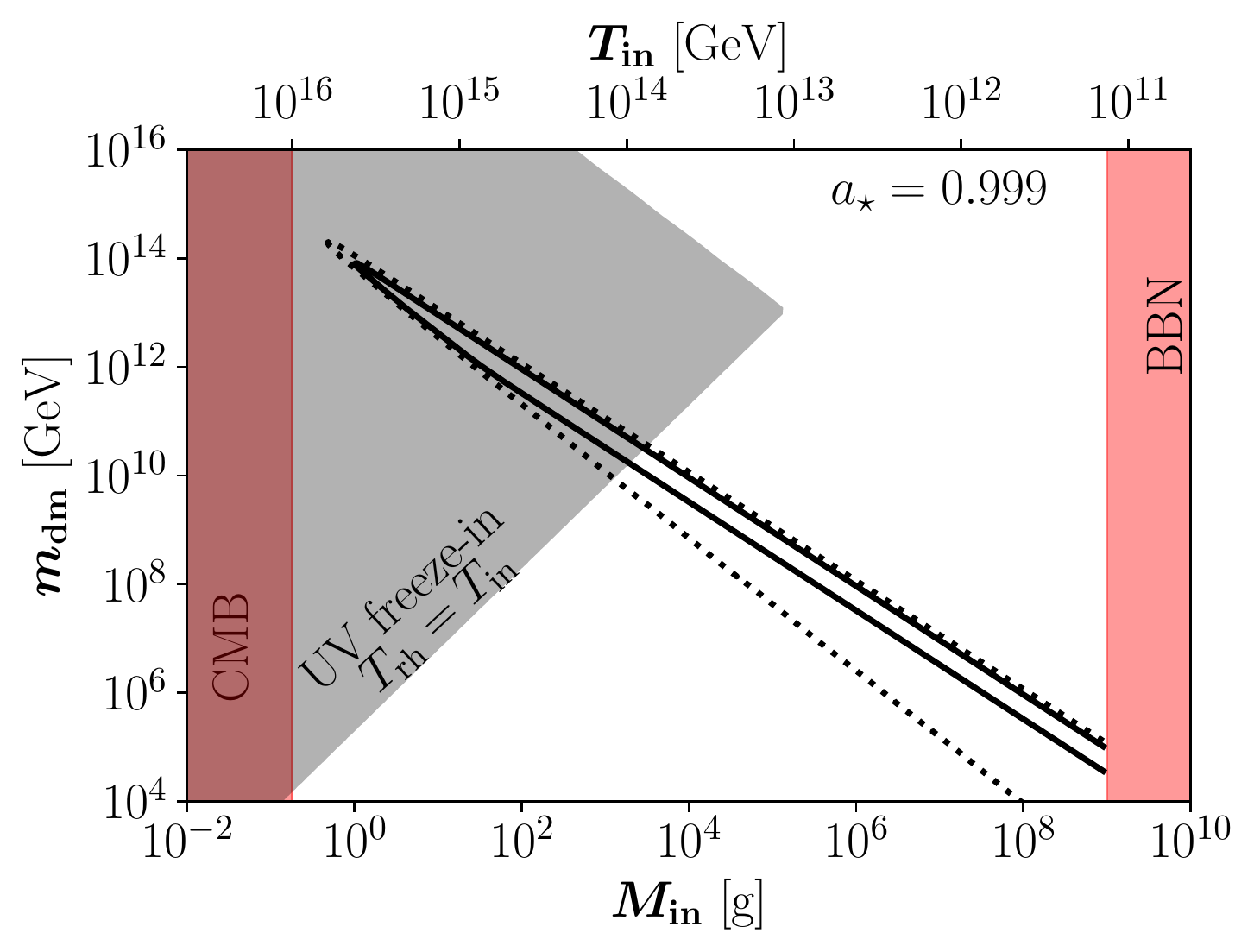}
    \includegraphics[scale=\sepf]{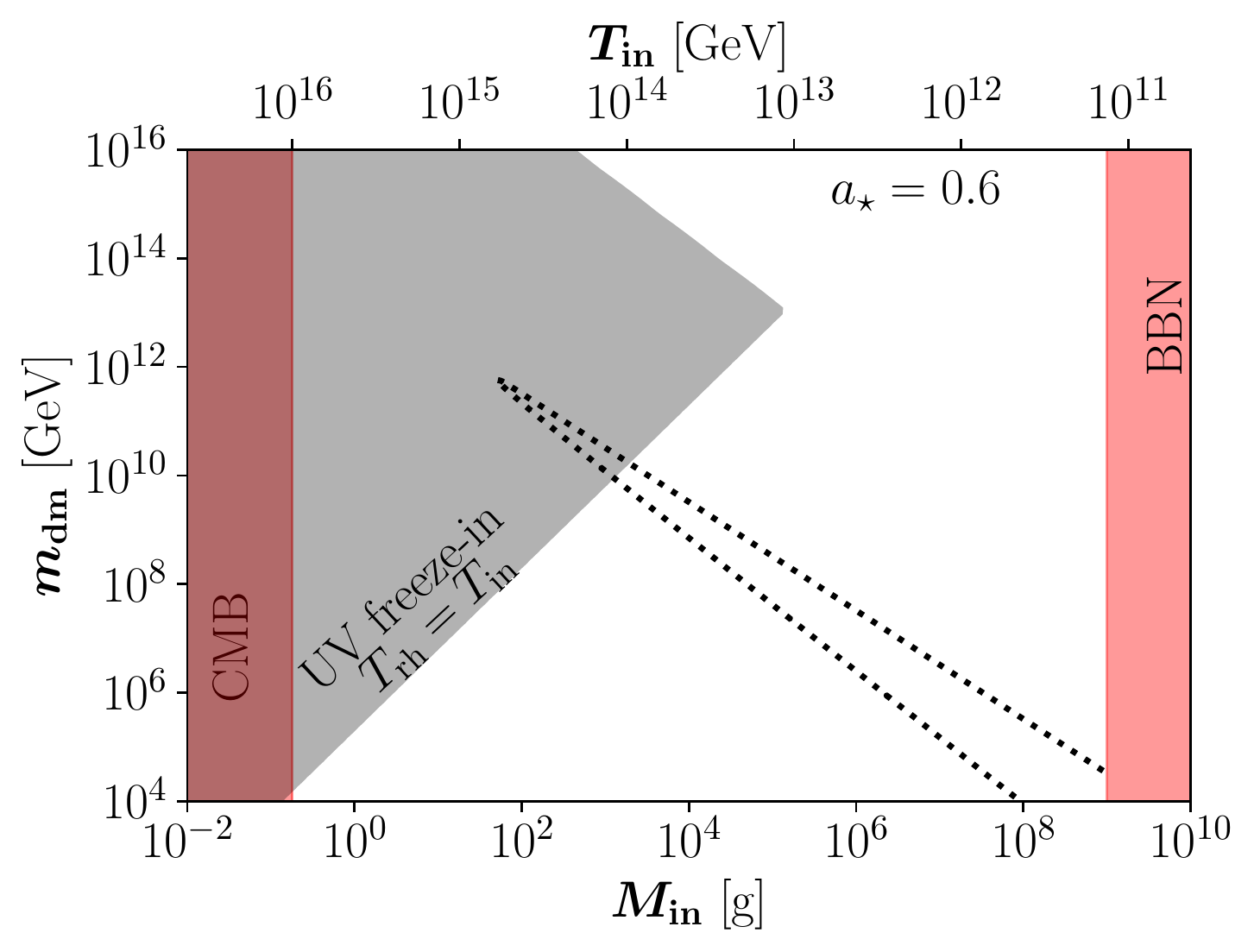}
    \caption{The solid and dotted black lines show the regions of the parameter space where ratio of the DM relic abundance with and without the inclusion of superradiance is 15 and 5, respectively.
    The left and right panels correspond to $a_\star = 0.999$ and $a_\star = 0.6$.
    In the gray area the universe could be overclosed due to the gravitational freeze-in, unless there is a large entropy dilution.
    The red regimes are excluded by different observables described in the text.
    }
	\label{fig:UV_sr}
\end{figure} 
The relative increase of the DM relic abundance due to superradiance can also be seen in Fig.~\ref{fig:UV_sr}.
The solid and dotted black lines show the regions of the parameter space where ratio of the DM relic abundances with and without the inclusion of superradiance is 15 and 5, respectively.
The left and right panels correspond to $a_\star = 0.999$ and $a_\star = 0.6$.
As expected, the maximum boost occurs for high values of the PBH spin and values of $\alpha \sim \mathcal{O}(1)$.
Additionally, the red bands are excluded by the CMB or BBN data.
For $a_\star = 0.999$, superradiance increases the relic abundance by at most a factor of $\sim 20$, for a large region of the parameter space, where the conditions are met.
For initial PBH masses $\Min\lesssim 10~{\rm g}$ and DM masses $\mdm\gtrsim 10^{12}~{\rm GeV}$, the region where there is an enhancement due to superradiance shrinks, since the depletion of angular momentum via Hawking radiation is fast enough so that the instability does not have time to develop.
In the case of heavier PBHs and lighter DM, superradiance can increase the relic abundance even if the gravitational coupling is small.
For example, for $\Min = 10^7$~g and $\mdm = 1.5\times 10^5$~GeV (corresponding to $\alpha \simeq 6\times 10^{-3}$), the relic abundance is enhanced by a factor $\sim 7$.
Furthermore, we observe that the region where the increase in the relic abundance is $\gtrsim 5$ grows for larger initial PBH masses.
In this case, PBHs have longer lifetimes that allow for the development of the instability, although at a lower rate since $\alpha\lesssim 0.1$.
As expected, for PBHs rotating with a smaller angular momentum, the increment on the final relic abundance is reduced.
For $a_\star = 0.6$ (cf. the right panel of Fig.~\ref{fig:UV_sr}), we find an increase by a factor of at most $\sim 7$.
Moreover, we observe that for initial PBH masses smaller than $\sim 70$~g, the superradiant instability does not develop (as implied by Eq.~\eqref{eq:Mines}), and the final relic abundance comes only from Hawking evaporation.

\section{Gravitational UV Freeze-in} \label{sec:uv}
Independently from the PBH evaporation, there is an irreducible DM production channel which is particularly efficient in the gray region in Fig.~\ref{fig:UV_sr} (i.e. large values for  the DM mass and the reheating temperature), and corresponds to the gravitational UV freeze-in~\cite{Garny:2015sjg, Tang:2017hvq, Garny:2017kha, Bernal:2018qlk}.
DM can be generated by 2-to-2 annihilations of SM particles, mediated by the exchange of massless gravitons in the $s$ channel.
We emphasize that this gravitational production mechanism is unavoidable due to universal couplings between the metric and the energy-momentum tensor $\sim g_{\mu \nu}\, h^{\mu \nu}$.

The evolution of the DM number density $n$ from gravitational freeze-in is given by the Boltzmann equation
\begin{equation} \label{eq:BEG}
    \frac{dn}{dt} + 3\, H\, n = \gamma\,,
\end{equation}
where $\gamma$ corresponds to the DM production rate density, given by~\cite{Garny:2015sjg, Tang:2017hvq, Bernal:2018qlk}
\begin{equation}
    \gamma(T) = \delta\, \frac{T^8}{M_P^4}\,,
\end{equation}
where $\delta \simeq 1.1 \times 10^{-3}$ for scalar DM.
Light DM particles (lighter than the reheating temperature) are mainly produced after the end of reheating. The DM yield from gravitational freeze-in $Y \equiv n(T)/s(T)$, with entropy density $s(T) = \frac{2\pi^2}{45} \gss T^3$, is obtained by integrating Eq.~\eqref{eq:BEG},
\begin{equation} \label{Y01}
	Y_0 = \frac{45\, \delta}{2\pi^3 \gss} \sqrt{\frac{10}{\gs}} \left(\frac{\Trh}{M_P}\right)^3, \qquad \text{ for } \mdm \ll \Trh\,.
\end{equation}
It is interesting to note that away from the instantaneous decay approximation, the DM yield is typically only boosted by a small factor of order $\mathcal{O}(1)$~\cite{Garcia:2017tuj, Bernal:2019mhf, Barman:2022tzk}.

Alternatively, in the case where DM is heavier than the reheating temperature (but still lighter than the highest temperature reached by the SM thermal bath~\cite{Giudice:2000ex}), it could not be produced after but during the reheating era.
Equation~\eqref{eq:BEG} therefore yields
\begin{equation}\label{Y02}
	Y_0 = \frac{45\, \delta}{2\pi^3 \gss} \sqrt{\frac{10}{\gs}} \frac{\Trh^7}{M_P^3\, \mdm^4}\,, \qquad \text{ for } \mdm \gg \Trh\,.
\end{equation}

We would like to stress that once produced by the UV freeze-in gravitational mechanism, the DM could suffer from a dilution due to the PBH evaporation, in the case where they manage to dominate the total energy density of the universe (i.e., if $\beta > \beta_c$).
The total entropy dilution is given by~\cite{Bernal:2021yyb}
\begin{equation}
    \frac{S(\Tin)}{S(\Tev)} \simeq
    \begin{dcases}
        1 & \text{ for } \beta < \beta_c\,,\\
        \frac{\Tev}{\Teq} \simeq \left(\frac{\gs}{640}\right)^\frac14 \left(\frac{M_P}{\Min}\right)^\frac32 \frac{M_P}{\beta\, \Tin} & \text{ for } \beta > \beta_c\,,
    \end{dcases}
\end{equation}
with the entropy defined as $S(T) = a^3 s(T)$.
To reproduce the observed DM relic density, one requires 
\begin{equation}
    \mdm\, Y_0\, \frac{S(\Tin)}{S(\Tev)} = \Omega_\text{dm} h^2 \, \frac{1}{s_0}\,\frac{\rho_c}{h^2} \simeq 4.3 \times 10^{-10}~\text{GeV},
\end{equation}
with $\rho_c \simeq 1.05 \times 10^{-5} \, h^2$~GeV/cm$^3$ being the critical energy density, $s_0\simeq 2.69 \times 10^3$~cm$^{-3}$ the entropy density at present~\cite{ParticleDataGroup:2020ssz}, and $\Omega_\text{dm} h^2 \simeq 0.12$ the observed DM relic abundance~\cite{Planck:2018vyg}.

The regions where the gravitational UV freeze-in is effective and can reproduce and even overshoot the whole observed DM abundance are overlaid in gray in Fig.~\ref{fig:UV_sr}.
Here, two assumptions were made: $i)$ as PBHs are assumed to be formed in a radiation-dominated universe, $\Trh \geq \Tin$.
Here we take $\Trh \geq \Tin$.
And $ii)$ no entropy dilution was considered, which can be achieved if $\beta < \beta_c$.

\section{Results} \label{sec:full}
\begin{figure}
    \def\sepf{0.51}
	\centering
    \includegraphics[scale=\sepf]{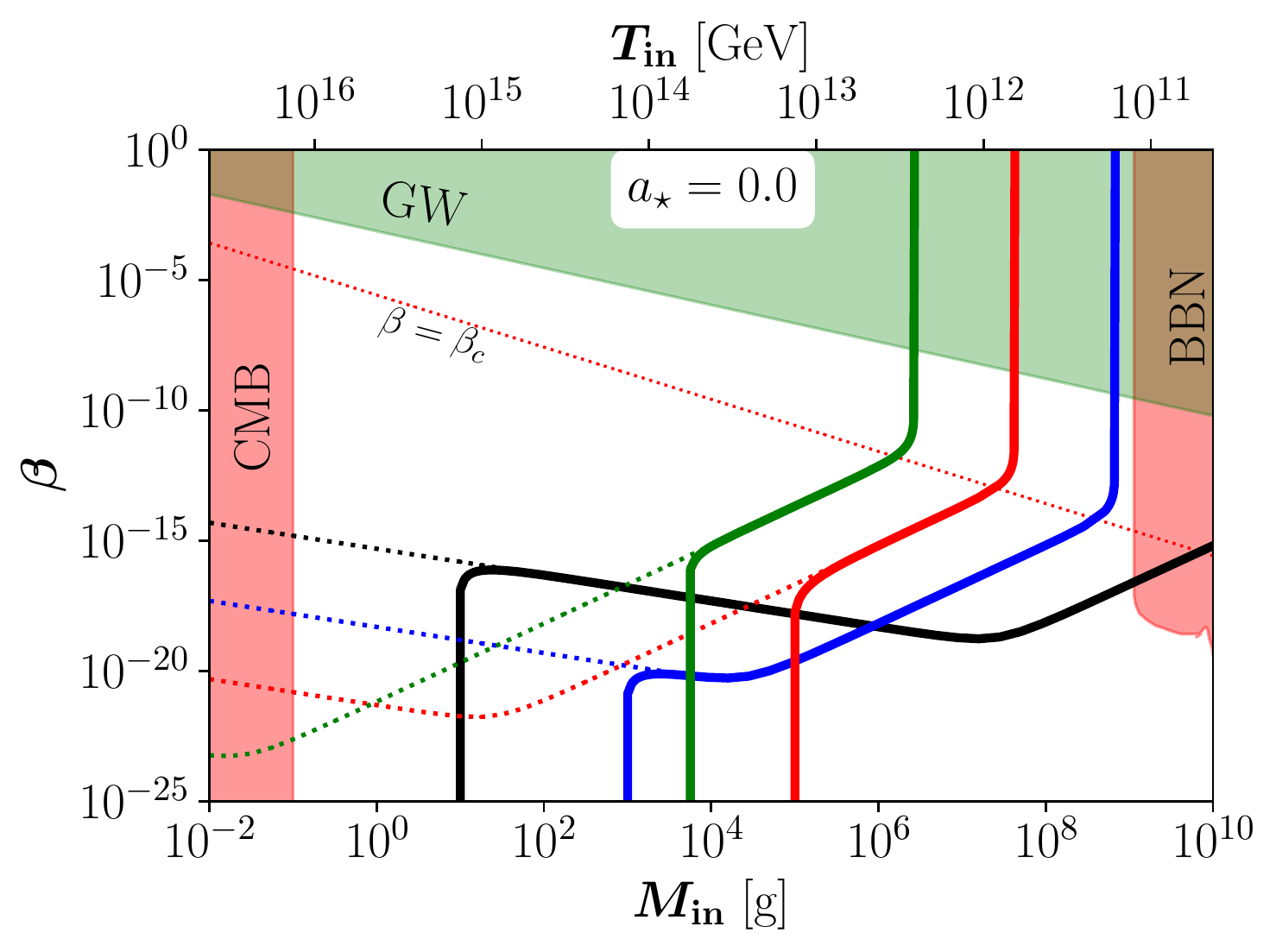}
    \includegraphics[scale=\sepf]{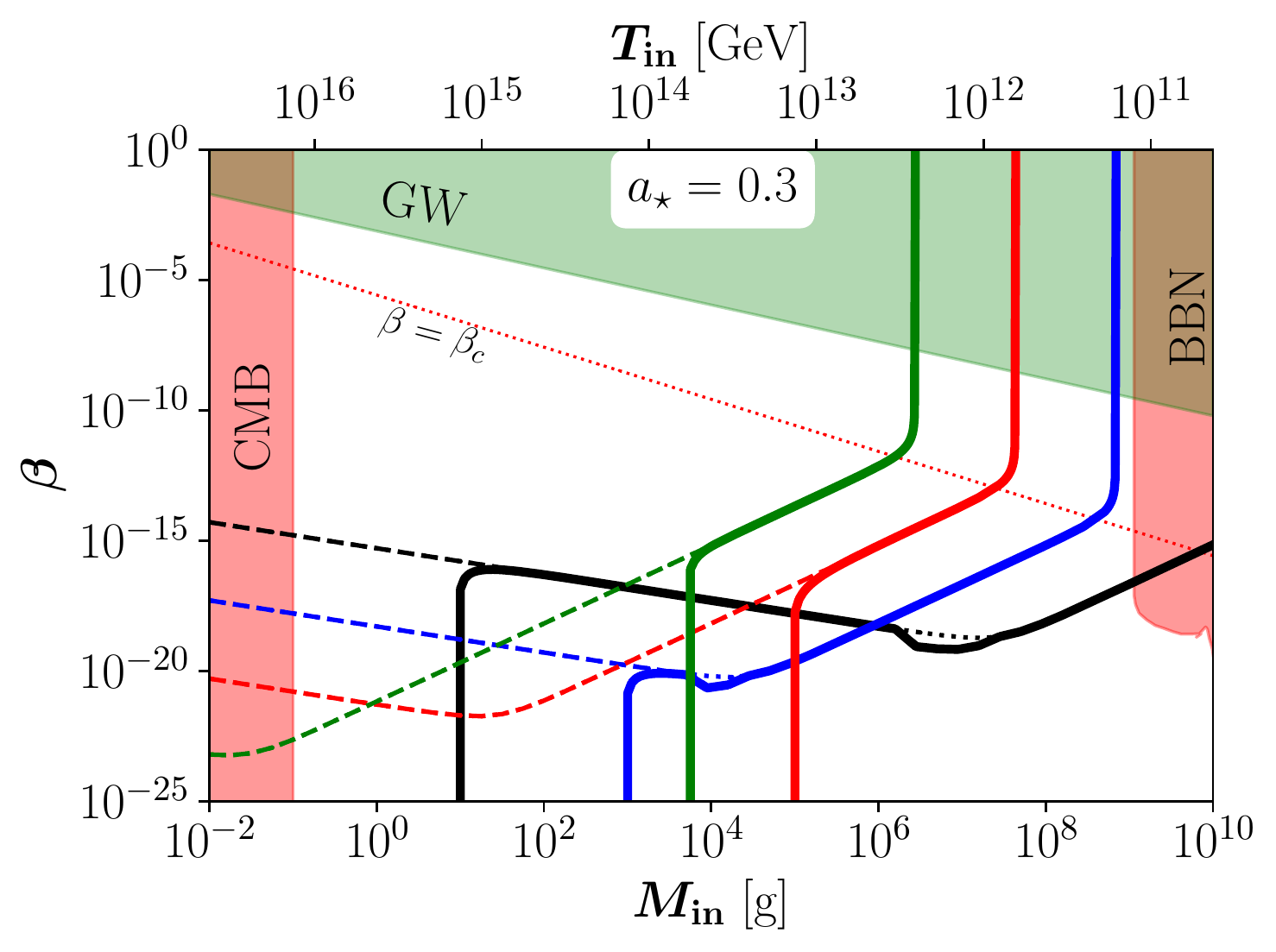}
    \includegraphics[scale=\sepf]{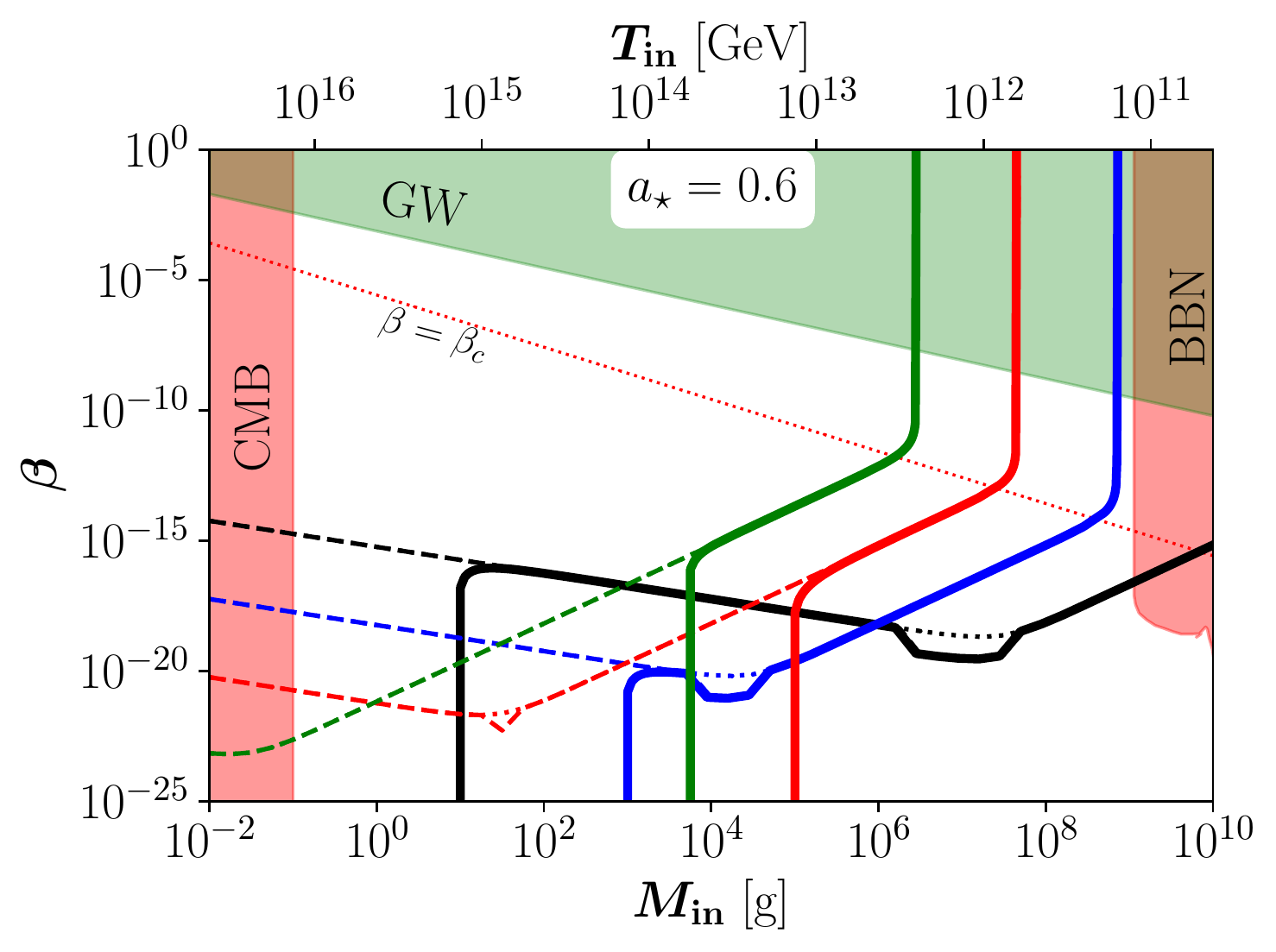}
    \includegraphics[scale=\sepf]{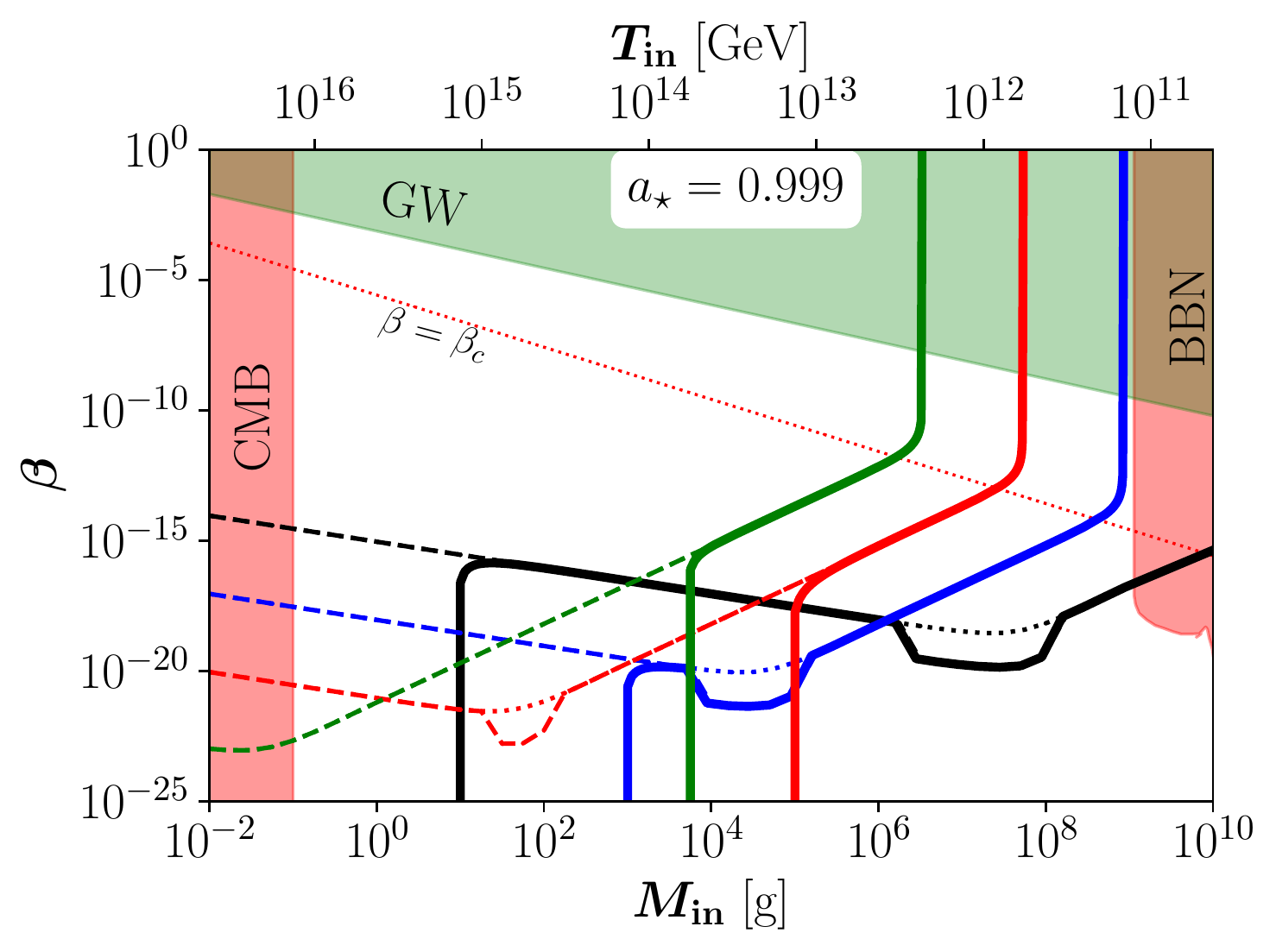}
    \caption{Parameter space that reproduces the whole observed DM abundance, for Hawking radiation only (dotted), Hawking radiation and superradiance (dashed), and Hawking radiation, superradiance and gravitational freeze-in (solid).
    The colors correspond to different DM masses $\mdm = 10^6$~GeV (black), $10^9$~GeV (blue), $10^{12}$~GeV (red) and $10^{15}$~GeV (green).
    The different panel correspond to $a_\star = 0.999$, 0.6, 0.3 and 0.
    The shaded areas are excluded by different observables described in the text.
    We assumed that $\Trh = \Tin$.
    }
	\label{fig:Min-beta}
\end{figure}

In this section, we summarize the comprehensive parameter space by taking into account the interplay of the aforementioned three mechanisms.
Figure~\ref{fig:Min-beta} shows with thick colored lines the parameter space $[\Min,\, \beta]$ reproducing the entire observed DM density for different DM masses: $\mdm = 10^6$~GeV (black), $10^9$~GeV (blue), $10^{12}$~GeV (red) and $10^{15}$~GeV (green), and four initial values for the spin of PBH $a_\star = 0$ (top left panel), $a_\star = 0.3$ (top right panel), $a_\star = 0.6$ (lower left panel) and $a_\star = 0.999$ (lower right panel).
Different line styles correspond to Hawking radiation only (dotted), Hawking radiation and superradiance (dashed), and Hawking radiation, gravitational freeze-in plus superradiance (thick solid).
The shaded regions represent areas constrained by different observables: $\Min \lesssim 10^{-1}$~g, cf. Eq.~\eqref{eq:CMB}, is disfavored by CMB, while large values for $\beta$ lead to modifications of BBN predictions due to the energy density stored in GWs, cf. Eq.~\eqref{eq:GW}.
For the case of the BBN, we present the bound obtained by studying in detail the evolution of the proton-to-neutron ratio and primordial abundances of light elements, including the evaporation of PBH, taken from Ref.~\cite{Carr:2009jm}.
Finally, the dotted line ($\beta = \beta_c$) shows the transition between radiation (lower part) and PBH-dominated eras (upper part).

Figure~\ref{fig:Min-beta} can be understood as follows.
First, the dotted lines correspond to the case where DM is only produced by Hawking radiation, and therefore this requirement shows three different slopes: $i)$ If PBHs dominated the energy density of the universe (for $\beta \gg \beta_c$), the DM yield is independent of $\beta$, and therefore the lines are vertical.
However, a dependence of $\beta$ appears if, during the entire lifetime of BH, the universe was always radiation dominated (for $\beta \ll \beta_c$).
$ii)$ If DM is lighter than the initial BH temperature $\beta \propto \Tin$, while in the opposite case $iii)$ $\beta \propto \Tin^{-3}$~\cite{Bernal:2020kse, Bernal:2020ili, Bernal:2021akf}.
In the present case where DM self-interactions are not efficient, DM has to be heavier than $\mathcal{O}(1)$~MeV in order not to be hot erasing seeds for structure formation~\cite{Masina:2020xhk, Bernal:2020kse, Baldes:2020nuv}, and can be as heavy as $M_P$~\cite{Chung:1999ve, Giudice:1999fb}.%
\footnote{Note that we are not considering a BH evaporation process stopping at $\Tbh\sim M_P$, with the associated production of Planck mass relics~\cite{MacGibbon:1987my, Barrow:1992hq, Carr:1994ar, Dolgov:2000ht, Baumann:2007yr, Hooper:2019gtx}.}

However, let us remember that such light PBHs are assumed to be created after the end of the reheating process, when the universe was SM-radiation dominated, at a temperature $\Tin \leq \Trh$.
The gravitational freeze-in production has a strong dependence on the reheating temperature, so that it can be minimized by taking $\Tin = \Trh$, which is the assumption used hereafter.
In that sense, this analysis is {\it conservative}, as we assume minimal possible gravitational freeze-in DM production.
In Fig.~\ref{fig:Min-beta} the dashed lines show the impact on the DM relic abundance once Hawking radiation and gravitational UV freeze-in are simultaneously taken into account.
The total abundance of DM can be produced by UV freeze-in for $\Min \lesssim 10^5$~ g and large values of DM masses, $\mdm \gtrsim 10^4$~GeV, cf. Fig.~\ref{fig:UV_sr}.
Therefore, a strong reduction in the $\beta$ parameter is needed to avoid DM overproduction.

Finally, the solid lines in Fig.~\ref{fig:Min-beta} take into account the three DM production mechanisms presented previously: Hawking radiation, gravitational freeze-in, and superradiance.
Superradiance is effective when the gravitational coupling $\alpha$ is close to unity.
As mentioned earlier and also shown in the right panel of Fig.~\ref{fig:solution}, there is a boost effect in the final DM number density once superradiance gets triggered.
This must be balanced with a reduction of $\beta$ in the regime when superradiant instability is efficient.
Since the superradiance rate $\Gsr \propto a_\star$, cf. Eq.~\eqref{eq:Gsr}, it is expected that the larger $a_\star$, the greater the reduction in $\beta$, which would be zero in the case $a_\star =0$ (upper left panel of Fig.~\ref{fig:Min-beta}).

\section{Conclusions} \label{sec:conclusion}
When the Compton wave length of a particle is comparable to the size of a spinning black hole (BH), energy can be efficiently transferred to the particle, leading to an amplification of occupation number in the ergoregion. For astrophysical BHs with mass $\Mbh \geq M_{\odot}$, it is possible to form clouds around BHs, which could give rise to some observational signatures; this has been utilized  as a potential new avenue to constrain ultralight particles beyond the standard model of particle physics~\cite{Brito:2015oca}. 

In this work, complementary to the astrophysical BHs as widely discussed in literature, we pay particular attention to the phenomenology of the superradiant instability with {\it primordial black holes} (PBHs) lighter than $\sim 10^9$~g. In particular, we investigate the impact of superradiance on the generation of {\it heavy {\rm ($\mdm \gtrsim 1$ TeV)} dark matter} (DM) particles in the early Universe. To this end we minimally set up a framework with the inevitable scenario where DM is generated pure gravitationally via either UV freeze-in or Hawking emission. Due to the exponential enhancement effect, we find that in the regime where superradiance manifests, the required BH abundance $\beta$ to yield the correct DM relics is reduced. 

\begin{figure}[t!]
    \def\sepf{0.9}
	\centering
    \includegraphics[scale=\sepf]{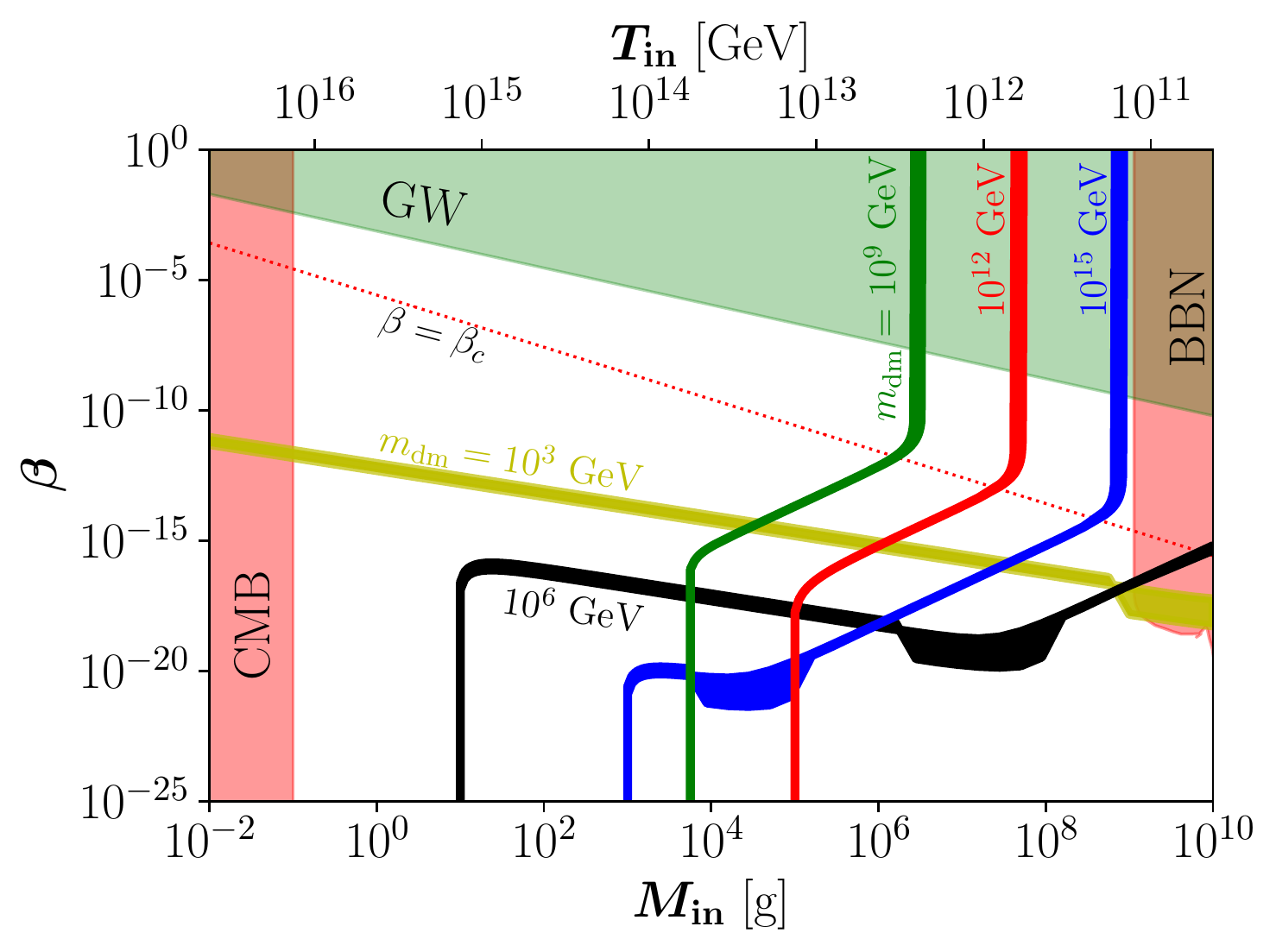}
    \caption{Parameter space that reproduces the whole observed DM abundance, taking the three contributions: Hawking radiation, superradiance and gravitational freeze-in.
    The colors correspond to different DM masses $\mdm = 10^3$~GeV (orange), $10^6$~GeV (black), $10^9$~GeV (blue), $10^{12}$~GeV (red) and $10^{15}$~GeV (green).
    The thickness of the bands corresponds to the different PBH spins: $0 \leq a_\star \leq 0.999$.
    The shaded areas are excluded by different observables described in the text.
    }
	\label{fig:Min-betaALL}
\end{figure}

Figure~\ref{fig:Min-betaALL} summarizes our main conclusions, where we show the parameter space that reproduces the whole observed DM abundance, taking into account all the gravitational production channels described previously (Hawking radiation, superradiance and gravitational freeze-in). The thickness of the bands brackets all possible values of the PBH spin: $0 \leq a_\star < 1$.
The color code is the same used before, i.e. $\mdm = 10^3$~GeV (orange), $10^6$~GeV (black), $10^9$~GeV (blue), $10^{12}$~GeV (red) and $10^{15}$~GeV (green).
We find that for a close-to-maximally rotating PBHs ($a_\star=0.999$) the relic abundance is increased by a factor of $\sim 20$, when the gravitational coupling between the DM and the PBH is $\sim{\cal O}(1)$.
For lower values of $a_\star$, the superradiant effect decreases and the observed relic abundance can be explained by purely Hawking evaporation.

\section*{Acknowledgments}
The authors would like to thank Manuel Drees for discussions and comments on the draft. 
We also would like to thank Jamie Mc Donald for his comments on the first version of this manuscript.
NB received funding from the Patrimonio Autónomo - Fondo Nacional de Financiamiento para la Ciencia, la Tecnología y la Innovación Francisco José de Caldas (MinCiencias - Colombia) grants 80740-465-2020 and 80740-492-2021.
NB is also funded by the Spanish FEDER/MCIU-AEI under grant FPA2017-84543-P.
This project has received funding/support from the European Union's Horizon 2020 research and innovation program under the Marie Skłodowska-Curie grant agreement No 860881-HIDDeN. 
YFPG would like to thank the Instituto de Fisica Teorica (IFT UAM-CSIC) in Madrid for support via the Centro de Excelencia Severo Ochoa Program under Grant CEX2020-001007-S, during the Extended Workshop ``Neutrino Theories'', where part of this work was completed.
This work has made use of the Hamilton HPC Service of Durham University. 

\appendix
\section{Massive Scalar Fields around Kerr Black Holes} \label{sec:appendix}
We revisit the solution of a massive scalar $\phi$ in the vicinity of a Kerr BH and further present the approximation of the superradiance rate $\Gsr$. We closely follow Ref.~\cite{Ferraz:2020zgi} where $G=\hbar =c=1$ is used, however we will restore the units at the end.  The Klein-Gordon equation for a massive scalar $\phi$ with mass $\mu$ in a Kerr metric, defined in Eq.~\eqref{eq:kerr}, is given by
\begin{equation}\label{eq:KG}
    \left(g_{\mu \nu} \nabla^\mu \nabla^\nu - \mu^2 \right) \phi = 0\,.
\end{equation}
Separating variables in a standard manner,
\begin{equation}
    \phi (t,r,\theta, \varphi) = \sum_{\omega, n,l. m} e^{-i\omega t + im \varphi} S_{lm}(\theta) R_{nlm}(r)\,,
\end{equation}
then one has  
\begin{align} \label{eq:EoM_S}
    &\frac{1}{\sin \theta} \partial_{\theta}\left(\sin \theta\, \partial_{\theta} S\right)+\left[a^2_\star r_g^{2}\left(\omega^{2}-\mu^{2}\right) \cos^{2} \theta-\frac{m^{2}}{\sin ^{2} \theta}+\lambda\right] S = 0\,, \\\label{eq:EoM_R}
    &\Delta \partial_{r}\left(\partial_{r} R\right)-\Delta\left[\mu^{2} r^{2}+a^2_\star r_g^{2} \omega^{2}-2 \omega \,m \,a_\star r_g\, r+\left(\omega\left(r^{2}+ a^2_\star r_g^{2}\right)-m\, a_\star r_g\right)^{2}+\lambda\right] R = 0\,,
\end{align}
where $\lambda$ denotes a separation constant, which is the eigenvalue of Eq.~\eqref{eq:EoM_S}.
Here our main interest is the solution of the radial component $R$, Eq.~\eqref{eq:EoM_R}, with a particular focus on the behavior of the frequency $\omega$ therein. Note that if the imaginary part $\mathfrak{Im} (\omega) < 0$, the wave would decay exponentially.  However, the amplitude of the field grows exponentially if $\mathfrak{Im} (\omega) > 0$.

Generically, there are no analytical solutions for Eq.~\eqref{eq:EoM_R}. 
However, we can consider the limit $\mu M \ll 1$, where $\lambda \simeq l(l+1)$~\cite{Detweiler:1980uk}. 
With a dimensionless coordinate $x = (r-r_+)/r_+$, it turns out that there are analytical solutions in two limit regimes with  near-horizon region $ x \ll l/\bar{\omega} $ and far region  $x\gg 1$, where we have defined $\bar{\omega} = r_+\, \omega$. 
One can then match the two solutions to obtain the field behavior in the overlapping region, $1 \ll x \ll  l/\bar{\omega} $, which includes the ergosphere.
There, an exponential growing mode appears if $\mathfrak{Im} (\omega)$ is positive.  This matching procedure leads to a constraint among the frequency $\omega$, BH mass $M$, as well as the quantum numbers, from which one can solve for $\mathfrak{Im} (\omega)$~\cite{Ferraz:2020zgi} 
\begin{equation}
    \mathfrak{Im} (\omega M )=-\frac{1}{2} A_{n l}\left(\frac{\omega^{\prime} M}{\tau}\right)(\mu M)^{4 l+5}\left(\frac{r_{+}-r_{-}}{r_{+}+r_{-}}\right)^{2 l+1},
\end{equation}
where $\omega^{\prime} = (2-\tau) (\omega -m \Omega)$, $\tau = (r_+ -r_-)/r_+$ and
\begin{equation}
    A_{n l}=\left(\frac{l !}{(2 l+1) !(2 l) !}\right)^{2} \frac{(l+n) !}{(n-l-1) !} \frac{4^{2 l+2}}{n^{2 l+4}} \prod_{k=1}^{l}\left[k^{2}+16\left(\frac{\omega^{\prime} M}{\tau}\right)^{2}\right].
\end{equation}
For the mode of interest with $n=2$ and $m = l = 1$, one has
\begin{equation} \label{eq:omegaM}
    A_{21} = \frac{1}{6} \left[1 + 16 \left(\frac{M (a_\star - 2 \omega\, r_+)}{2 r_+ \sqrt{1-a_\star^2}}\right)^2\right]\,,
\end{equation}
and
\begin{align} \label{eq:omegaM2}
    \mathfrak{Im} (\omega M ) & =-\frac{1}{2} A_{2 1}\left(\frac{\omega^{\prime} M}{\tau}\right)(\mu M)^{9}\left(\frac{r_{+}-r_{-}}{r_{+}+r_{-}}\right)^{3}\nonumber \\
    &=  \frac{1}{24} (\mu M)^{9} (a_\star - 2 \omega\, r_+)  \frac{  (1 - a_\star^2)}{\left(1 + \sqrt{1-a_\star^2}\right)} + \mathcal{O} \left( \alpha  - \frac{1}{2} \right)^3.
\end{align}
Note that in the Schwarzschild limit with $a_\star \to 0$, one has $r_+ \to 0$, and then $\mathfrak{Im} (\omega M )$ vanishes, as expected. 
At leading order $\omega \sim \mu$, one has
\begin{equation}
    a_\star - 2 \omega r_+ \simeq a_\star - 2 \mu r_+
    \simeq -2\left(1 + \sqrt{1-a_\star^2}\right) \left( \alpha  - \frac{1}{2} \right),
\end{equation} 
where we have considered $\Omega \sim 1/(2\, r_g)$, cf. Eq.~\eqref{Eq:Omega}. 
Due to the superradiance condition, $\alpha < 1/2$, higher order terms of  $\left( \alpha  - 1/2 \right)$ in Eq.~\eqref{eq:omegaM2} can be omitted. 
For $n=2$ and $m = l =1$, one has
\begin{equation}
    \mathfrak{Im} (\omega M )  \simeq \frac{1}{24} (\mu M)^{9} (a_\star - 2 \omega\, r_+)\,,
\end{equation}
where $(1 - a_\star^2)/\left(1 + \sqrt{1-a_\star^2}\right) \sim \mathcal{O}(1)$ has been assumed.\footnote{We note that this approximation fails for large values of $a_\star$.}
Defining the superradiance growth rate as $\Gsr \equiv \mathfrak{Im} (\omega)$, one  has
\begin{equation}
    \Gsr = \frac{\mu}{24}  (\mu M)^{8} (a_\star - 2 \mu \, r_+)\,,
\end{equation}
which corresponds to Eq.~\eqref{eq:Gsr} once restoring $G^{-1}=8\pi M_P^2$.

\bibliographystyle{JHEP}
\bibliography{biblio}

\end{document}